\documentclass[aps,prb,reprint,superscriptaddress]{revtex4-2}

\usepackage{etoolbox}
\usepackage{graphicx,amssymb,bm,amsfonts,amsmath,graphics,color,epstopdf}
\usepackage{mathrsfs}
\usepackage{upgreek}
\usepackage[bookmarks=false,pdfstartview=FitH,colorlinks=true,citecolor=blue,linkcolor=blue]{hyperref}
\usepackage[version=3]{mhchem}
\usepackage{lipsum, babel}
\usepackage[super]{nth}
\usepackage{overpic}
\usepackage{bbm}
\AtBeginEnvironment{align}{\setcounter{subeqn}{0}}
\newcounter{subeqn} %

\begin{document}

\title{Impact of strain and dark states on spectroscopic measurements of\\ silicon-vacancy centers in diamond}

\author{Tommy Chin}
\thanks{These authors contributed equally to this work.}
\affiliation{Department of Physics and Astronomy, San Jos\'e State University, San Jose, CA 95192, USA}
\affiliation{Institute for Gravitation and the Cosmos and Department of Physics, The Pennsylvania State University, University Park, PA 16802, USA}
\author{Kesav V.\ Narayan}
\thanks{These authors contributed equally to this work.}
\affiliation{Department of Physics and Astronomy, San Jos\'e State University, San Jose, CA 95192, USA}
\author{Imran Bashir}
\thanks{These authors contributed equally to this work.}
\affiliation{Department of Physics and Astronomy, San Jos\'e State University, San Jose, CA 95192, USA}
\author{Kelsey M.\ Bates}
\affiliation{Department of Physics, University of Michigan, Ann Arbor, MI 48109, USA}
\author{Liam G.\ Stanton}
\affiliation{Department of Mathematics and Statistics, San Jos\'e State University, San Jose, CA 95192, USA}
\author{Ehsan Khatami}
\affiliation{Department of Physics and Astronomy, San Jos\'e State University, San Jose, CA 95192, USA}
\author{Christopher L.\ Smallwood}
\email[Email:\ ]{christopher.smallwood@sjsu.edu}
\affiliation{Department of Physics and Astronomy, San Jos\'e State University, San Jose, CA 95192, USA}
\date {\today}

\begin{abstract}
Negatively charged silicon-vacancy (\ce{SiV^-}) centers in diamond offer an attractive platform for the development of many forms of quantum technology. However, questions remain in connection to how large ensembles of \ce{SiV^-} centers behave in concert. Here, we develop a computational model designed to simulate recent experiments where optical multidimensional coherent spectroscopy (MDCS) was used to examine a high-concentration sample of \ce{SiV^-} centers in diamond, revealing significant variations in spectral signature depending on the detection scheme. Simulation results reveal that strain effects are highly random in this system, with a characteristic axial strain of $2.8 \times 10^{-4}$ and a shear strain of $3.5 \times 10^{-5}$. They suggest in addition that highly strained centers (with values exceeding $1.5 \times 10^{-5}$) may become significantly decoupled from optical emission. The results have implications for the use of \ce{SiV^-} centers as quantum sensors.
\end{abstract}

\maketitle

\section{Introduction}
\label{sec:intro}

Recent years have witnessed an increasing need to characterize and optimize the properties of quantum light sources for use in sensing, cryptography, computing, and communication \cite{Rodgers2021}. Within this realm, the negatively charged silicon-vacancy (\ce{SiV^-}) center in diamond has emerged as especially promising. For example, the system exhibits high-quality single-photon emitter properties \cite{Aharonovich2014,Becker2017}, and researchers have developed techniques to enhance and manipulate optical absorption and emission \cite{Evans2018, Sohn2018,Meesala2018,Koppenhofer2023,Ngan2023,Assumpcao2023,Lindner2025}. Proof-of-principle demonstrations have been reported using \ce{SiV^-} centers in applications ranging from blind quantum computing \cite{Wei2025} to entanglement-assisted optical interferometry \cite{Stas2026} to localized material strain sensing \cite{Meesala2018,Knauer2020,Bates2021}.

Device impacts aside, color centers in diamond offer opportunities for studying basic research questions, including what happens when color centers are placed in close proximity to each another or in highly strained environments \cite{Bradac2017,Venkatesh2018,Kucsko2018,Lindner2018,Smallwood2021,Angell2024}. Questions of this sort have been relevant, for example, in a recent study where multidimensional coherent spectroscopy (MDCS) was used to examine a high-density sample of \ce{SiV^-} centers in chemical-vapor-deposition (CVD) grown single-crystal diamond, comparing heterodyne and photoluminescence detection schemes \cite{Smallwood2021}. The former scheme monitors coherent emission, and it is sensitive to radiative and non-radiative electronic states alike due to the rapid nature of the emission process. The latter scheme is preferentially sensitive to strongly radiatively-coupled states. Surprisingly, the authors found that heterodyne-detected measurements exhibited more than 60 times as much spectral inhomogeneity as did the photoluminescence-detected measurements. To explain their results, the authors developed a model positing strain-dependent coupling between electromagnetically accessible ``bright" states and an unobserved ``dark" state. The model was successful in qualitative terms, but it remains to be seen what happens after more realistic parameters are incorporated. Such improvements might inform, for example, the mechanisms of strain-impacted processes in the \ce{SiV^-} center system more generally.

To that end, in this Article, we develop a quantitative model of strain effects in the \ce{SiV^-} system studied in Ref.~\cite{Smallwood2021}, which sets constraints on both magnitude and orientation propertes. We find that strain effects are likely to be maximally random, with little-to-no systematic correlation between tensor elements. We estimate an axial strain variation in the system of $2.8\times10^{-4}$ and a shear strain variation of $3.5\times10^{-5}$. To reproduce the photoluminescence detection data, strain values in excess of $1.5\times10^{-5}$ are expected to become significantly attenuated in terms of optical coupling. Overall, the results demonstrate the power of MDCS and the utility of comparing different kinds of collection schemes \cite{Maly2018,Kunsel2019}. They also serve as an invitation to consider the possibility that optically detectable states of the \ce{SiV^-} center system may not be the only ones present or relevant.


\section{Background and Formalism}
\label{sec:background}

\begin{figure}[tb]\centering\includegraphics[width=3.375in]{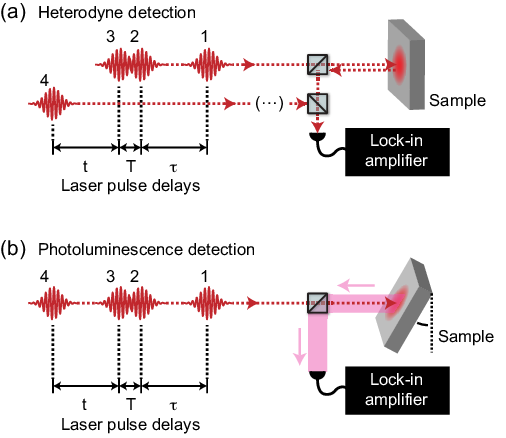}
\caption{Schematic illustration of MDCS measurements of \ce{SiV^-} centers in diamond as reported in Ref.~\cite{Smallwood2021}. A series of time-delayed laser pulses excites the sample, producing a four-wave mixing response that can be read out as coherent emission of light or as an incoherent photoluminescence beatnote.
{\bf(a)} Heterodyne detection scheme for measuring coherent emission. 
{\bf(b)} Photoluminescence detection scheme.}
\label{schematic}
\end{figure}

Figure~\ref{schematic} illustrates the experimental layout relevant to the simulations presented in this work. A sequence of laser pulses separated by time delays $\tau$, $T$, and $t$ is directed toward the \ce{SiV^-} center sample, generating a nonlinear material response. This response emerges either as a coherently emitted four-wave mixing signal [Fig.~\ref{schematic}(a)] or as a beatnote in the incoherently emitted photoluminescence [Fig.~\ref{schematic}(b)]. The signal is recorded as a function of the inter-pulse time delays, and the results are then Fourier-transformed to produce a spectrum unfolded across a multidimensional frequency domain. MDCS can thus be understood as an optical-frequency analog of two-dimensional nuclear magnetic resonance (2D-NMR), allowing access to coherent coupling signatures and many-body effects in materials that one-dimensional measurements cannot resolve. Beyond this capability, pulse orders and collection schemes can be readily engineered in MDCS experiments to produce rephasing ({\it i.e.}, Hahn-echo) signal responses that in turn allow the practitioner to see through 
the often deleterious effects of inhomogeneous spectral broadening.

We conduct simulations using density-matrix perturbation theory within a Bloch-model framework and in the rotating-wave approximation~\cite{Boyd,Li}. To simplify the analysis, we take both the impulsive and Markovian limits. The impulsive limit treats laser pulse envelopes as Dirac delta functions, and the Markovian limit treats decoherence phenomena and population decay as exponential. We restrict our attention to single-quantum rephasing spectra and retain only the signal pathways with ${\bf k}_\text{sig} = -{\bf k}_1 + {\bf k}_2 + {\bf k}_3$ and $\nu_\text{sig} = -\nu_1 + \nu_2 + \nu_3$.

\begin{figure}[tb]\centering\includegraphics[width=3.375in]{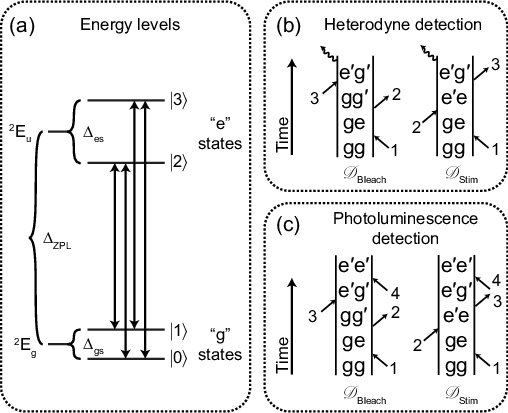}
\caption{
{\bf(a)} Energy level structure for the \ce{SiV^-} center system in diamond.
{\bf(b)} Rephasing double-sided Feynman diagrams for heterodyne detection.
{\bf(c)} Rephasing double-sided Feynman diagrams for photoluminescence detection.
}
\label{energylevels}
\end{figure}

The \ce{SiV^-} center electronic level structure to be studied emerges from a $^2E_g$ ground state and a $^2E_u$ excited state. Spin-orbit coupling splits each of these into a pair of substates, producing a four-level system as depicted in Fig.~\ref{energylevels}(a). At a temperature of 10--15 K, thermal fluctuations populate both ground substates, so all four absorption lines appear.

Accounting for each of these constraints, the MDCS signal can be expressed \cite{Li} as the sum
\begin{equation}
    S(t,T,\tau) = \sum_\alpha {\mathscr D}_\alpha(t,T,\tau), \label{firstsum}
\end{equation}
where the elements ${\mathscr D}_\alpha(t,T,\tau)$ correspond to products of decaying exponentials encoding the \mbox{first-,} \mbox{second-,} and third-order light-matter interactions as
\begin{equation}
{\mathscr D}_\alpha = A_\alpha
\underbrace{\Theta(t) e^{-i \Omega_{e'g'} t} }_{3^{\text{rd}}\text{-order}}
\underbrace{\Theta(T) e^{-i \Omega_{mn} T} }_{2^{\text{nd}}\text{-order}}
\underbrace{\Theta(\tau) e^{-i \Omega_{ge} \tau} }_{1^{\text{st}}\text{-order}}. \label{timedomain}
\end{equation}
In these expressions, $\alpha=\{e',g',m,n,g,e\}$ runs over the various density matrix indices relevant to the process, $A_\alpha$ is a constant, and $\Theta(x)$ is the Heaviside step function. The parameters
\begin{equation}
\Omega_{ab} \equiv \omega_{ab} - i \gamma_{ab} \quad \text{and} \quad \omega_{ab} \equiv \frac{E_a - E_b}{\hbar}
\label{complexfrequency}
\end{equation}
combine the dephasing rates $\gamma_{ab}$ with the resonant frequencies $\omega_{ab}$ between energy levels $E_a$ and $E_b$.

Double-sided Feynman diagrams \cite{Boyd,Li} representing the terms described by Eq.~\eqref{timedomain} are shown for heterodyne detection in Fig.~\ref{energylevels}(b) and for photoluminescence detection in Fig.~\ref{energylevels}(c)\@. In each of the two schemes, the left diagram represents $\Omega_{mn} = \Omega_{gg'}$, corresponding to ground-state bleaching processes and related pathways, and the right diagram represents $\Omega_{mn} = \Omega_{e'e}$, corresponding to stimulated-emission processes and related pathways. No other values of $\Omega_{mn}$ are allowed. Thus, for heterodyne-detected and photoluminescence-detected MDCS schemes alike, for each of the two different diagram types, there are 16 different permutations [the variables $g$, $g'$, $e$, and $e'$ can take on one of two values, depending on the ground or excited state represented in Fig.~\ref{energylevels}(a)]. The only measurable difference between detection schemes in this system is the collection efficiency as governed by $A_\alpha$. As discussed later on in Section \ref{sec:PL}, this collection efficiency factor can be strain-dependent in principle. However, the quasi-two-level system relevant to \ce{SiV^-} centers is otherwise relatively simple to analyze, bypassing the excited-state absorption effects that might impede the interpretation of results in something like a three-state ladder system \cite{Gregoire2017,Maly2018,Kunsel2019,Kalaee2019}.

Fourier-transforming Eq.~\eqref{timedomain} with respect to the first- and third-order time delays $\tau$ and $t$ while holding the second-order time delay fixed at $T=0$ turns each double-sided Feynman diagram term into a complex-valued two-dimensional Lorentzian,
\begin{equation}
\label{lorentzian}
{\mathscr D}_\alpha(\nu_t,\nu_\tau) =
\frac{A_\alpha}{4\pi^2} \left(\frac{i}{\nu_t - \tilde\Omega_{e'g'}}\right) \left(\frac{i}{\nu_\tau - \tilde\Omega_{ge}}\right),
\end{equation}
where $\tilde\Omega_{ab} \equiv \Omega_{ab}/2\pi = \nu_{ab} - i\gamma$ with $\nu_{ab} = \omega_{ab}/2\pi$ and $\gamma = \gamma_{ab}/2\pi$ being the resonant frequency and dephasing rate in Hz. We take a single dephasing rate for all transitions, hence the missing subscript. 

\begin{figure}[tb]\centering\includegraphics[width=3.375in]{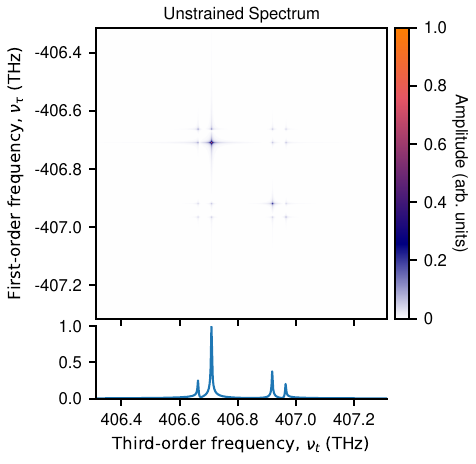}
\caption{Unstrained \ce{SiV^-} rephasing spectrum. The 16 Lorentzian peaks arise from all permutations of ground- and excited-state indices, with amplitudes taken from Ref.~\cite{Smallwood2021}. The lower panel shows the projection onto $\nu_t$.
\label{lorentzians}
}
\end{figure}

We can see, in consequence, that the peak position of Eq.~\eqref{lorentzian} depends on the poles $\tilde\Omega_{ge}$ and $\tilde\Omega_{e'g'}$, set by the first- and third-order coherences. The left and right diagrams of Figs.~\ref{energylevels}(b) and \ref{energylevels}(c) therefore generate a grid of spectral peaks as displayed in Fig.~\ref{lorentzians}. Writing ground- and excited-state splittings as $\Delta_\text{gs}$ and $\Delta_\text{es}$ and the zero-phonon line frequency as $\Delta_\text{ZPL}$ [see Fig.~\ref{energylevels}(a)], the pole positions take on the form
\begin{equation}
    \label{eq:omega}
    \tilde\Omega_{ab} = \Delta_\text{ZPL} \pm \tfrac{1}{2}\Delta_\text{gs}
    \pm \tfrac{1}{2}\Delta_\text{es} - i\gamma,
\end{equation}
thereby codifying the various ground-state/excited-state transitions into a unified equation. 

Strain enters the parameters in Eq.~\eqref{eq:omega} as $\Delta_\text{ZPL}(\boldsymbol{\epsilon})$, $\Delta_\text{gs}(\boldsymbol{\epsilon})$, and $\Delta_\text{es}(\boldsymbol{\epsilon})$, where $\boldsymbol{\epsilon} \equiv (\epsilon_{xx}, \epsilon_{yy}, \epsilon_{zz}, \epsilon_{xy}, \epsilon_{yz}, \epsilon_{zx})$ collects the strain components in the \ce{SiV^-} internal frame, which is the frame set by each center's own symmetry axes. Their dependence follows the treatment derived in Ref.~\cite{Hughes1967} and extended to the \ce{SiV^-} center system in Ref.~\cite{Meesala2018}, giving
\begin{align}
&\begin{array}{ll}
\Delta_\text{ZPL} = 
        \Delta_{\text{ZPL},0} &+ \left( t_{\parallel,\text{es}} - t_{\parallel,\text{gs}} \right) \epsilon_{zz} \\[6pt]
        &+ \left( t_{\perp,\text{es}} - t_{\perp,\text{gs}} \right) \left( \epsilon_{xx} + \epsilon_{yy} \right),
    \end{array} \label{Meesala1} \\[10pt]
&\Delta_\text{gs} = \sqrt{
    \begin{array}{ll}
        \lambda^2_{\text{SO},\text{gs}} 
        &+ 4 \left[ d_\text{gs} \left( \epsilon_{xx} - \epsilon_{yy} \right) + f_\text{gs} \epsilon_{zx} \right]^2 \\[6pt] 
        &+ 4 \left[2 d_\text{gs} \epsilon_{xy} - f_\text{gs} \epsilon_{yz} \right]^2
    \end{array}} \label{Meesala1b}, \\[10pt]
&\Delta_\text{es} = \sqrt{
    \begin{array}{ll}
        \lambda^2_{\text{SO},\text{es}} 
        &+ 4 \left[ d_\text{es} \left( \epsilon_{xx} - \epsilon_{yy} \right) + f_\text{es} \epsilon_{zx} \right]^2 \\[6pt]
        &+ 4 \left[2 d_\text{es} \epsilon_{xy} - f_\text{es} \epsilon_{yz} \right]^2
    \end{array}}, \label{Meesala2}
\end{align}
with $\Delta_{\text{ZPL},0}$ being the unstrained zero-phonon line frequency, $\lambda_{\text{SO},\text{gs}}$ and $\lambda_{\text{SO},\text{es}}$ being the spin-orbit coupling parameters, and $t_{\parallel}$, $t_{\perp}$, $d$, and $f$ being susceptibility coefficients determining the effect of strain values on electronic energy-level values. The placement of $\epsilon_{zx}$ and $\epsilon_{yz}$ in Eqs.~\eqref{Meesala1b} and \eqref{Meesala2} follows from diagonalizing the strain Hamiltonian of Ref.~\cite{Meesala2018} and corrects an erroneous transposition of those two components reported later on in that work.

\section{Results: Heterodyne-detected Spectra}
\label{sec:HDS}

Having laid out the formalism for incorporating strain into \ce{SiV^-} center spectral signatures in generalized terms, we proceed with more detailed simulations corresponding to experimental data. The measured sample of Ref.~\cite{Smallwood2021} contains many \ce{SiV^-} centers, each experiencing a different local strain environment. We model this signal by taking a sum over $N$ Monte Carlo-sampled strain configurations,
\begin{equation}
    S_\text{tot} \propto \sum_{j=1}^{N} S(\boldsymbol{\epsilon}_j),
    \label{eq:ensemble_sum}
\end{equation}
Each $\boldsymbol{\epsilon}_j$ is an independently sampled strain configuration, drawn from whichever distribution the model in question specifies. $N$ is large enough that Eq.~\eqref{eq:ensemble_sum} closely approximates the true sum over the sample's much larger population of physical centers.

We accumulate the sum by Monte Carlo sampling, accumulating $8 \times 10^4$ randomly sampled strain configurations on a $2000 \times 2000$ frequency grid. We determine $\Delta_{\text{ZPL},0}$ and $\gamma$ from Ref.~\cite{Smallwood2021}, with $\Delta_{\text{ZPL},0}$ being calculated from the mean of the four measured transitions reported there, and the spin-orbit coupling parameters and strain susceptibility coefficients from Ref.~\cite{Meesala2018}, as listed in Table~\ref{parameters}.

\begin{table}[tb]
\centering
\caption{Strained \ce{SiV^-} center coupling parameters extracted from Refs.\ \cite{Meesala2018} and \cite{Smallwood2021}.}
\begin{ruledtabular}
\begin{tabular}{l r@{\,$\times$\!\!\!\!\!\!\!\!\!\!\!\!\!\!\!\!\!\!\!\!\!\!}l l}
Symbol & \multicolumn{2}{l}{\quad Value} & Unit \\
\hline
$\Delta_{\text{ZPL},0}$ & \multicolumn{2}{l}{\,\, 406.814} & THz \\
$\gamma$ & $1.33$ & $10^{-3}$ & THz \\
$\lambda_{\text{SO},\text{gs}}$ & $46$ & $10^{-3}$ & THz \\
$\lambda_{\text{SO},\text{es}}$ & $255$ & $10^{-3}$ & THz \\
$t_{\parallel,\text{es}} - t_{\parallel,\text{gs}}$ & $-1.7$ & $10^{3}$ & THz/strain \\
$t_{\perp,\text{es}} - t_{\perp,\text{gs}}$ & $7.8$ & $10^{1}$ & THz/strain \\
$d_\text{gs}$ & $1.3$ & $10^{3}$ & THz/strain \\
$d_\text{es}$ & $1.8$ & $10^{3}$ & THz/strain \\
$f_\text{gs}$ & $-1.7$ & $10^{3}$ & THz/strain \\
$f_\text{es}$ & $-3.4$ & $10^{3}$ & THz/strain
\end{tabular}
\end{ruledtabular}
\label{parameters}  
\end{table}

\subsection{Fully Random Anisotropic Strain}

\begin{figure*}[tb]\centering
\includegraphics[width=7in]{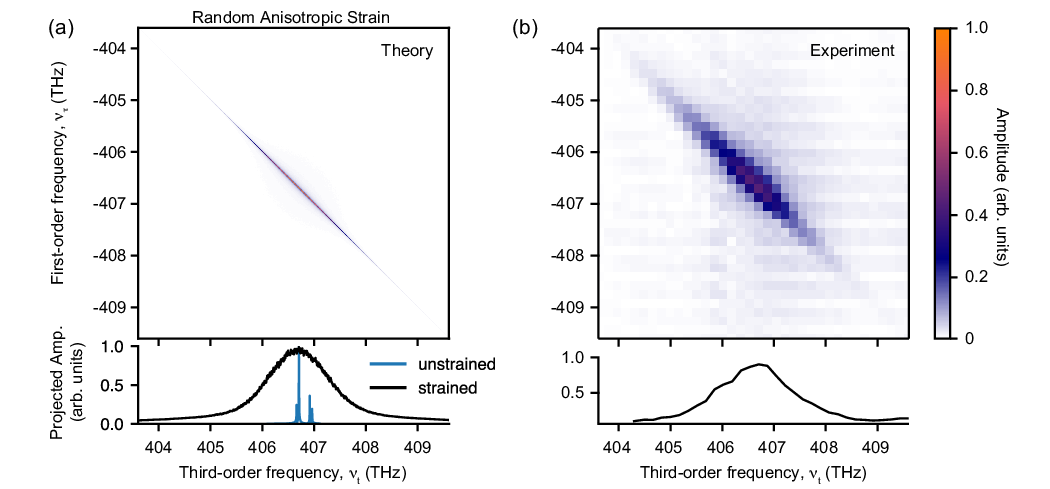}
\caption{Comparison between theory and experiment for strained heterodyne-detected MDCS measurements of \ce{SiV^-} centers in diamond. 
{\bf(a)} Simulated spectrum under circumstances of fully random anisotropic strain, with parameters as specified in the main text. The upper panel shows the two-dimensional amplitude $|S_\text{tot}(\nu_t, \nu_\tau)|$. The lower panel shows the projection of this amplitude onto the $\nu_t$ axis (black trace). The projection of the unstrained spectrum amplitude (blue trace) from Fig.~\ref{lorentzians} is also shown for comparison.
{\bf(b)} Experimental dataset for heterodyne-detected MDCS measurements of \ce{SiV^-} centers in diamond as reported in Ref.~\cite{Smallwood2021}.
\label{rand6}}
\end{figure*}

The most general model of strain effects in the \ce{SiV^-} center system varies all six strain components independently, reflecting the disorder expected in a densely implanted sample. Researchers also use this approach to describe the small residual strain present in nominally unstrained \ce{SiV^-} center ensembles~\cite{Assumpcao2023}. Defining the distribution in the \ce{SiV^-} internal frame of each center, we draw each component independently from a Gaussian distribution centered at zero, as illustrated in Fig.~\ref{rand6}(a) and compared to experimental data reported in Ref.~\cite{Smallwood2021} in Fig.~\ref{rand6}(b).

In the illustrated version of the simulation, $\sigma_\text{axial}$ sets the width of the axial strains $\epsilon_{ii}$, and $\sigma_\text{shear}$ sets the width of the shear strains $\epsilon_{i<j}$. The axial strains enter the spectrum through the zero-phonon line frequency $\Delta_\text{ZPL}(\boldsymbol{\epsilon})$ as indicated in Eqs.~\eqref{Meesala1}--\eqref{Meesala2}, and they translate the peaks along the diagonal line $\nu_\tau = -\nu_t$. Shear strains and the difference $\epsilon_{xx} - \epsilon_{yy}$, by contrast, modulate the splittings $\Delta_\text{gs}(\boldsymbol{\epsilon})$ and $\Delta_\text{es}(\boldsymbol{\epsilon})$ and redistribute the peak positions. We can see in this way that there is an advantage to assigning separate widths to the two kinds of strain values because of their independent impact on spectral features. More concretely, independent control over the two parameters lets the model reproduce a mild asymmetry in the measured projected lineshape [see Fig.~\ref{rand6}(b), lower panel], concave upward on its low-frequency side and concave downward on its high-frequency side. A single shared width could not have captured this. 

In addition to replicating lineshape asymmetry, we seek to reproduce the overall lineshape width. To accomplish this goal, we employ an empirical taper criterion: The two-dimensional amplitude spectrum $|S_\text{tot}(\nu_t,\nu_\tau)|$, smoothed to suppress Monte Carlo sampling noise, must fall below 5\% of its own peak value at the edge of the plotted frequency window. We apply this same criterion separately to each strain group here and to the corresponding width parameter for every other strain model in this section. We then adjust each value of $\sigma$ within the range the criterion allows to match the measured lineshape's asymmetry by eye. For fully random anisotropic strain, this constraint gives $\sigma_\text{axial} = 2.8\times10^{-4}$ and $\sigma_\text{shear} = 3.5\times10^{-5}$, producing the single diagonally elongated feature shown in Fig.~\ref{rand6}. Both the shape and extent of this feature match the heterodyne-detected spectrum measured in Ref.~\cite{Smallwood2021}, as well as the expected strain values based on implantation density estimates reported in that same work.

\subsection{Alternate Models}

\begin{figure*}[!ht]\centering
\includegraphics[width=6.1in]{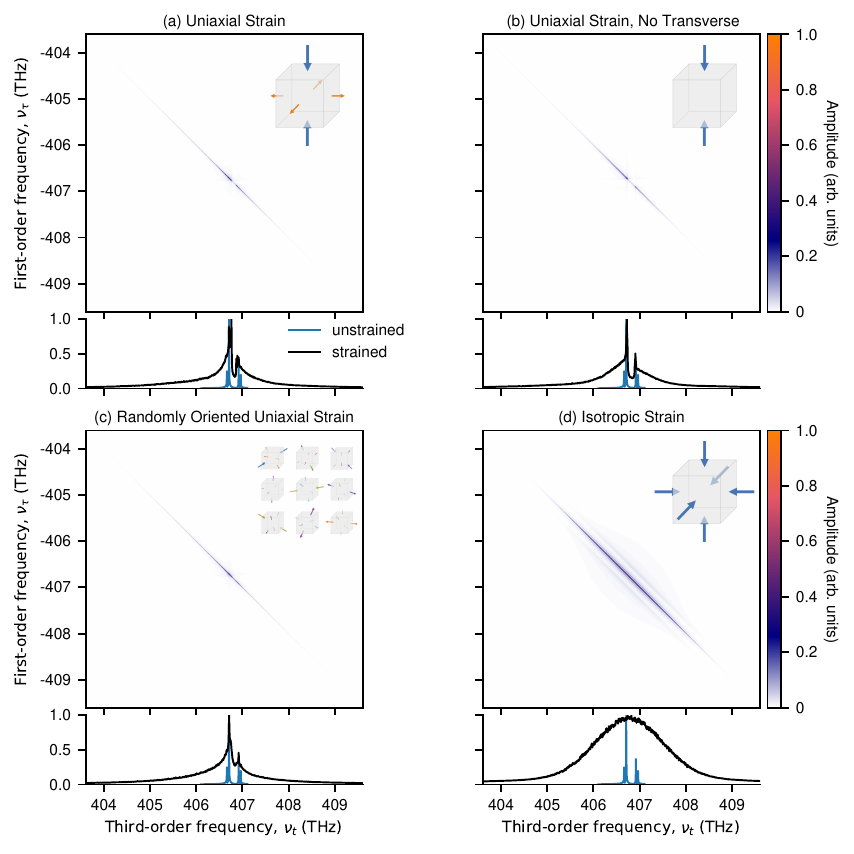}
\caption{Heterodyne-detected MDCS spectra for four alternative strain models. Each panel shows the two-dimensional amplitude spectrum above and its projection onto the $\nu_t$ axis below, with the unstrained reference overlaid in blue.
{\bf (a)}~Uniaxial strain.
{\bf (b)}~Uniaxial strain without transverse response, $p=0$.
{\bf (c)}~Randomly oriented uniaxial strain.
{\bf (d)}~Isotropic strain.
The cube schematic in the upper right corner of each panel illustrates the corresponding strain, with arrows indicating the strain directions. In~(c), several cubes at different orientations represent the orientational average. The arrows in panels (a), (b), and (d) depict the compressive sign only and do not represent the Gaussian ensemble, which spans both compression and expansion.
\label{otherheterodyne}}
\end{figure*}

Treating all six strain components as independent is the least structured assumption available. To understand which features of that model drive the observed broadening, we examine four structured models in which the strain follows a definite geometric form, with results illustrated in Fig.~\ref{otherheterodyne}. None of these models reproduces the measured spectrum as well as fully random anisotropic strain. Each nonetheless isolates the spectral signature of a particular kind of strain and builds intuition for how a given strain geometry appears in an MDCS plot.

The simplest structured model applies uniaxial strain along $\hat{y}=[110]$. This direction is normal to the sample surface, along which stress from fabrication and handling of the sample would predominantly act. Uniaxial stress compresses the lattice along $\hat{y}$ and, through the elastic response of the material, expands the lattice along the two transverse directions $\hat{x}=[1\bar{1}0]$ and $\hat{z}=[001]$. The uniaxial strain of magnitude $\epsilon_0$ along $\hat{y}$ takes the matrix form used below for the change of basis,
\begin{equation}\label{eq:uniDirStrain}
    \hat{\epsilon} = \epsilon_0
    \begin{pmatrix}
        -p & & \\
        & 1 & \\
        & & -p
    \end{pmatrix},
\end{equation}
where the Poisson ratio $p$ sets the magnitude of the transverse response along $\hat{x}$ and $\hat{z}$. We set $p = 0.2$, which is admittedly an overestimate for bulk single-crystal diamond~\cite{Klein1993} but is consistent with measurements of nanodiamonds~\cite{Mohr2014} and large enough to produce an observable spectral effect. The cube schematic in the upper right corner of Fig.~\ref{otherheterodyne}(a) illustrates this strain geometry with the compressive sign shown. The Monte Carlo average draws $\epsilon_0$ from a Gaussian centered at zero, so the ensemble spans both compression and expansion across a range of magnitudes.

\begin{figure}[tb]\centering
\includegraphics[width=1in]{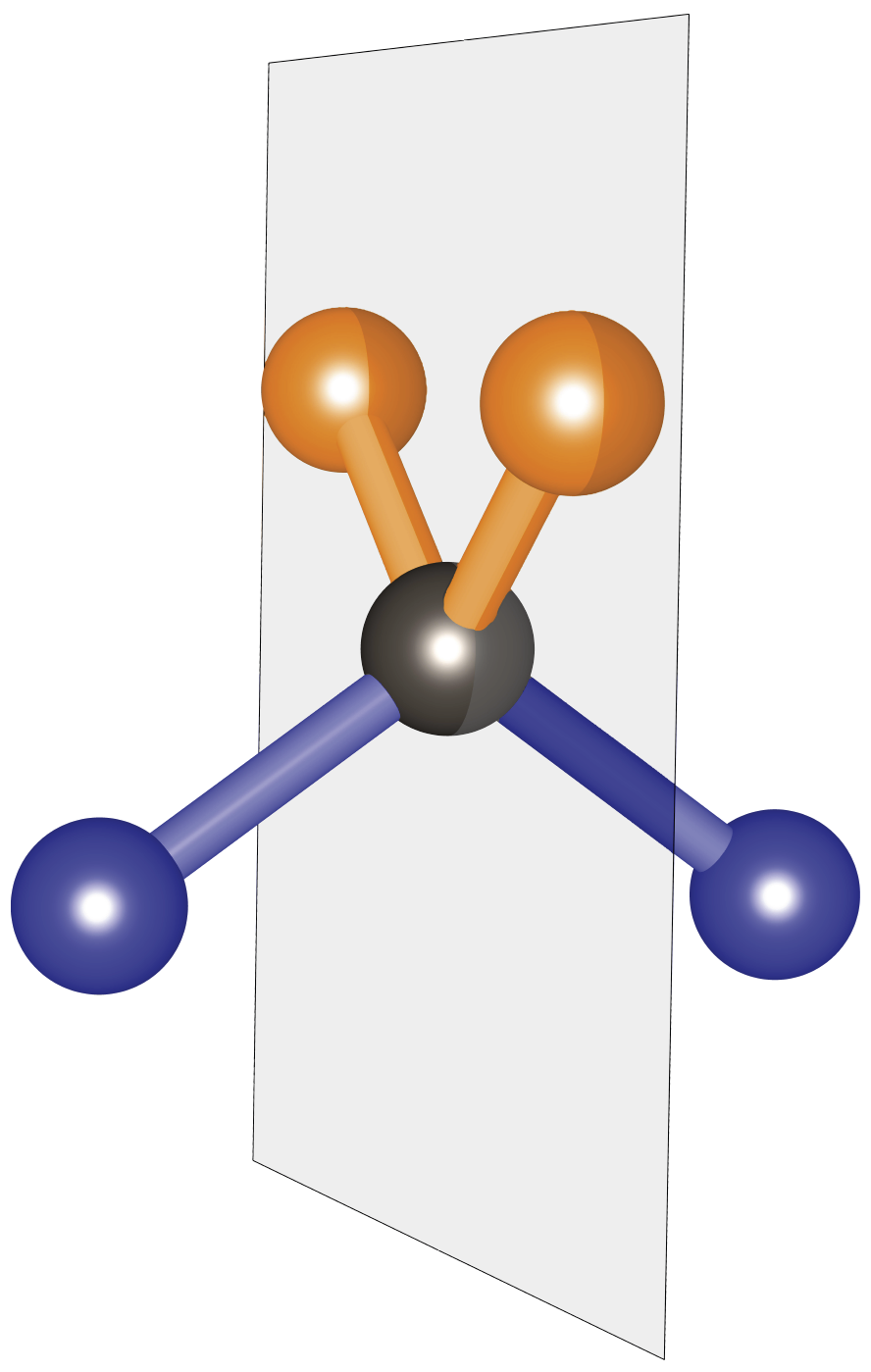}
\caption{Illustration of the in-plane \ce{SiV^-} orientation (orange) and the out-of-plane orientation (indigo). The plane shown is parallel to the sample surface.
\label{SiV}}
\end{figure}

Because we define this strain in the crystal frame while Eqs.~\eqref{Meesala1}--\eqref{Meesala2} hold in the \ce{SiV^-} internal frame~\cite{Meesala2018}, evaluating the signal requires a change of basis from the crystal frame into the internal frame of each \ce{SiV^-} center. The \ce{SiV^-} symmetry axis lies along one of the four $\langle 111\rangle$ directions, which split relative to the sample surface into two in-plane axes and two axes tilted out of the surface. Figure~\ref{SiV} illustrates these two orientations, color-coded orange and indigo respectively, and the change of basis proceeds separately for each. The fully random model required no such step, because we define the distribution there directly in each center's internal frame.

The in-plane \ce{SiV^-} symmetry axis sits at $\theta_0 = \arccos(1/\sqrt{3})$ from $\hat{z}$, so the change of basis transforms $\hat\epsilon$ by the rotation matrix $\hat{R}_y(\theta_0)$. Since $\hat{\epsilon}$ has equal diagonal components along $\hat{x}$ and $\hat{z}$, this transformation leaves the tensor unchanged, $\hat{\epsilon}_\text{in} = \hat{R}_y(\theta_0)\, \hat{\epsilon}\, \hat{R}_y^T(\theta_0) = \hat{\epsilon}$. Writing $\boldsymbol{\epsilon}_\text{in}$ for the vector formed from the six components of $\hat{\epsilon}_\text{in}$, we draw $\epsilon_0$ from a Gaussian distribution with standard deviation $\sigma$ and average $S(\boldsymbol{\epsilon}_\text{in})$ for the in-plane contribution. Each structured model below writes the strain as $\hat\epsilon = \epsilon_0 \hat{M}$, and a change of basis rotates $\hat{M}$ while leaving the scalar magnitude $\epsilon_0$ untouched, so $\sigma$ refers to the same standard deviation in any frame.

The out-of-plane \ce{SiV^-} symmetry axis tilts out of the sample surface, and the change of basis for this orientation composes a rotation by $\theta_0$ about $\hat{x}$ with a rotation by $-\pi/2$ about $\hat{z}$, giving
\begin{align}
    \hat{\epsilon}_\text{out}
    &= \hat{R}_z\!\left(-\tfrac{\pi}{2}\right) \hat{R}_x(\theta_0)\,
       \hat{\epsilon}\, \hat{R}_x^T(\theta_0)\, \hat{R}_z^T\!\left(-\tfrac{\pi}{2}\right)
       \notag \\
    &= \frac{\epsilon_0}{3}
    \begin{pmatrix}
        1 - 2p & & \sqrt{2}(1+p)\\
        & -3p & \\
        \sqrt{2}(1+p) & & 2-p
    \end{pmatrix}.
\end{align}
The off-diagonal components of $\hat\epsilon_\text{out}$, already expressed in the \ce{SiV^-} internal frame, introduce shear strain into the splittings $\Delta_\text{gs}(\boldsymbol{\epsilon})$ and $\Delta_\text{es}(\boldsymbol{\epsilon})$, a contribution absent in the in-plane case. Writing $\boldsymbol{\epsilon}_\text{out}$ for the vector formed from the six components of $\hat{\epsilon}_\text{out}$, we average $S(\boldsymbol{\epsilon}_\text{out})$ using the same randomly distributed $\epsilon_0$ as for the in-plane contribution to obtain the out-of-plane contribution. Splitting the ensemble sum of Eq.~\eqref{eq:ensemble_sum} into in-plane and out-of-plane sub-populations and evaluating each separately gives the total signal as $S_\text{tot} = S_\text{tot,in} + S_\text{tot,out}$. The taper criterion sets $\sigma = 9.0\times10^{-4}$, matching the value used for the simulation depicted in Fig.~\ref{rand6}, and gives the spectrum in Fig.~\ref{otherheterodyne}(a).

Because a single strain direction shifts and splits the four transition frequencies together rather than smearing them independently, the discrete peak structure survives the ensemble average. In the projection onto $\nu_t$, the ensemble average smears away the small gaps of width $\Delta_\text{gs}(\boldsymbol\epsilon)$ between the central and outer peaks. The larger gap $\Delta_\text{es}(\boldsymbol\epsilon) - \Delta_\text{gs}(\boldsymbol\epsilon)$ between the two central peaks survives instead as a pronounced dip, a structure that does not match the single broad feature reported in Ref.~\cite{Smallwood2021}.

Setting the Poisson ratio to $p = 0$ removes the transverse response. However, the out-of-plane population still carries nonzero shear strain, since the off-diagonal terms of $\hat\epsilon_\text{out}$ remain $\sqrt{2}$ under these circumstances. Such a model would align more closely with the accepted Poisson ratio for bulk single-crystal diamond of $p=0.0691$~\cite{Klein1993}. The taper criterion gives $\sigma = 9.0\times10^{-4}$, indistinguishable from the value required with the transverse response included, and produces the spectrum in Fig.~\ref{otherheterodyne}(b). The central feature is narrower than that seen in Fig.~\ref{otherheterodyne}(a), though both share the same qualitative structure, discrete peaks separated by a pronounced central dip that does not match the measured spectrum.

A third model incorporates randomly oriented uniaxial strain. In a disordered sample, the local strain axis varies from \ce{SiV^-} center to center. We model this by orienting the uniaxial strain tensor $\hat{\epsilon}$ of Eq.~\eqref{eq:uniDirStrain} uniformly over all directions and averaging the result. We parameterize each orientation by its strain axis $\hat{n}$, set by angles $\theta$ and $\phi$ about the $\hat{x}$- and $\hat{z}$-axes,
\begin{equation}
    \hat{\epsilon}(\epsilon_0, \hat{n}) = \hat{R}_z(\phi)\, \hat{R}_x(\theta)\,
    \hat{\epsilon}\, \hat{R}_x^T(\theta)\, \hat{R}_z^T(\phi).
\end{equation}

A change of basis into the \ce{SiV^-} internal frame followed by the orientational average gives the same distribution as the orientational average alone, so the average requires no separate change of basis. With $\epsilon_0$ drawn from a Gaussian with $\sigma = 8.0\times10^{-4}$ Using our taper criterion and the strain axis distributed uniformly over the sphere, we write $\boldsymbol{\epsilon}(\epsilon_0, \hat{n})$ for the vector formed from the six components of $\hat{\epsilon}(\epsilon_0, \hat{n})$. Averaging $S(\boldsymbol{\epsilon}(\epsilon_0, \hat{n}))$ over this distribution gives the spectrum shown in Fig.~\ref{otherheterodyne}(c). The orientational average partially fills the dip produced by the uniaxial models, but the discrete structure persists, and the spectrum does not recover the single broad feature of Fig.~\ref{rand6}.

A fourth model applies isotropic strain, $\hat{\epsilon} = \epsilon_0\, \hat{\mathbbm{1}}$, representing the effect of uniform hydrostatic pressure and compressing or expanding the lattice equally along every direction. Because the identity matrix is invariant under any change of basis, the ensemble average requires no change of basis into the \ce{SiV^-} internal frame. In contrast to the uniaxial models, isotropic strain carries no shear, and the difference $\epsilon_{xx} - \epsilon_{yy}$ vanishes identically for every orientation. In this case, we choose $\sigma = 4.5\times10^{-4}$ by the taper criterion, giving the spectrum in Fig.~\ref{otherheterodyne}(d). Isotropic strain enters only through the zero-phonon line frequency in Eq.~\eqref{Meesala1}, leaving the splittings $\Delta_\text{gs}(\boldsymbol\epsilon)$ and $\Delta_\text{es}(\boldsymbol\epsilon)$ at their unstrained values. Every peak therefore translates rigidly along the diagonal, and the spectrum preserves the discrete four-peak structure as the ensemble broadens, with the outer peaks surviving rather than washing out. The projection onto the $\nu_t$ axis washes into a single broad envelope resembling the fully random anisotropic result of Fig.~\ref{rand6}, but the two-dimensional spectrum retains the discrete structure that the measured heterodyne spectrum does not show~\cite{Smallwood2021}. Isotropic strain and fully random anisotropic strain are therefore distinguishable in the two-dimensional lineshape despite producing near-identical one-dimensional projections, illustrating the added diagnostic value of the full MDCS spectrum over a one-dimensional measurement.

Of the five models, only fully random anisotropic strain reproduces the measured lineshape. Nothing guaranteed this result in advance. The sample is single-crystal, for example, and therefore breaks rotational symmetry. Implantation also confined silicon ions to the diamond sample's upper layers, opening possibilities for unidirectional effects. Regardless, fully random anisotropic strain is also the only model that lets every strain component, including all three shear components, vary independently rather than fixing them by geometry or setting them to zero. The uniaxial models retain shear only for a fraction of centers and still show discrete structure. Isotropic strain has no shear at all and shows discrete structure even more sharply. Reproducing the measured broadening therefore requires shear strain acting independently across the full ensemble, not merely present in some orientations.

\section{Results: Photoluminescence-detected Spectra}
\label{sec:PL}

\begin{figure*}[tb]\centering
\includegraphics[width=7in]{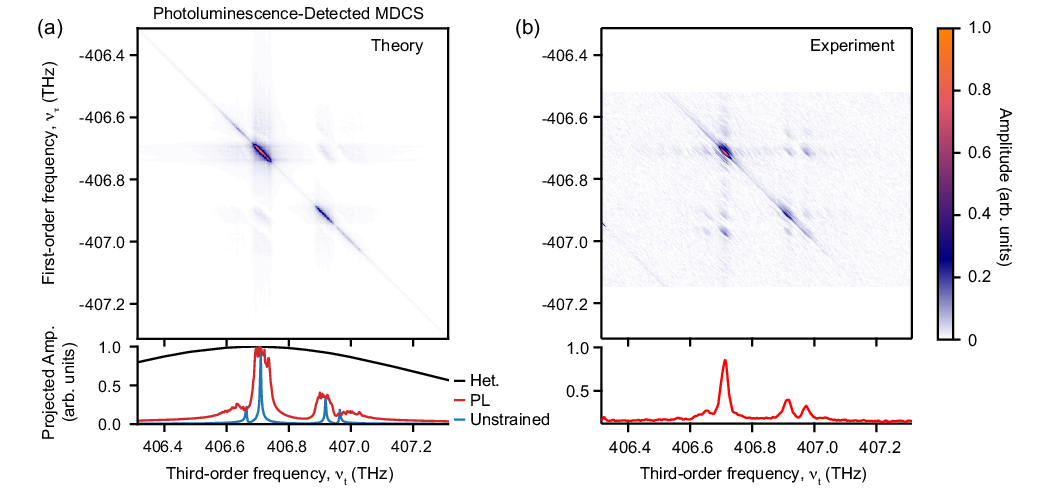}
\caption{Comparison between theory and experiment for strained photoluminescence-detected MDCS measurements of \ce{SiV^-} centers in diamond. 
{\bf(a)} Simulated spectrum. The upper panel shows the two-dimensional amplitude, and the lower panel shows the projection onto the $\nu_t$ axis (red trace, labeled ``PL"), as well as a heterodyne-detected reference spectrum (black trace, labeled ``Het.") and unstrained reference spectrum (blue trace), shown for comparison. 
{\bf(b)} Experimental dataset for photoluminescence-detected MDCS measurements of \ce{SiV^-} centers in diamond, as reported in Ref.~\cite{Smallwood2021}.
}
\label{spectrum_PRL}
\end{figure*}

Having modeled the possible effects of strain on reproducing heterodyne-detected spectra, we turn our attention to photoluminescence detection and how the differences between photoluminescence- and heterodyne-detected spectra might lead to additional subtleties and insights. As mentioned in Sections \ref{sec:intro} and \ref{sec:background}, heterodyne detection measures the third-order coherent response directly, independently of how the excited population subsequently decays. Photoluminescence detection instead responds to the fourth-order population that relaxes radiatively over much longer timescales, so a nonradiative channel competing with radiative decay suppresses the signal. We model this suppression through a branching ratio for each excited substate,
\begin{equation}
    B_{e}(\boldsymbol{\epsilon}) = \frac{\Gamma_{\text{rad},e'}}{\Gamma_{\text{rad},e'} + \Gamma_\text{nr}(\boldsymbol{\epsilon})},
    \label{branchingratio}
\end{equation}
defined identically for each of the two excited states. Here, $\Gamma_{\text{rad},e'}$ is the radiative decay rate from excited state $e'$, and $\Gamma_\text{nr}(\boldsymbol{\epsilon})$ is a strain-dependent nonradiative rate coupling that state to a dark state~\cite{Neu2012, Gali2013, Thiering2018}. This branching ratio follows the form derived in Ref.~\cite{Smallwood2021}. In the unstrained center, symmetry forbids the transition into the dark state, so $\Gamma_\text{nr} = 0$ and $B_{e'} = 1$. Strain breaks that symmetry and opens the nonradiative channel, suppressing the photoluminescence signal from strongly strained centers.

The single-center photoluminescence signal $S_\text{PL}(\boldsymbol{\epsilon})$ weights each pathway of $S(\boldsymbol{\epsilon})$ by the branching ratio of its terminating excited substate,
\begin{equation}
    S_\text{PL}(\boldsymbol{\epsilon}) = \sum_\alpha B_{\alpha}(\boldsymbol{\epsilon})\, {\mathscr D}_\alpha,
\end{equation}
where $B_\alpha$ denotes whichever of $B_{e'}$ corresponds to the excited substate pathway on which $\alpha$ terminates. Applying the same Monte Carlo sum as in Eq.~\eqref{eq:ensemble_sum} to $S_\text{PL}(\boldsymbol{\epsilon})$ over the fully random anisotropic strain distribution gives $S_\text{PL,tot}$.

We approximate $\Gamma_\text{nr}$ as a step function of the strain, which is zero when the largest strain component magnitude lies below a threshold of $1.5\times10^{-5}$ and a value much larger than $\Gamma_{\text{rad},e'}$ for strain values above that threshold. Resolving the filtered four-peak structure against the residual broadened background requires finer sampling than the heterodyne spectra needed, so we accumulate $10^8$ randomly sampled strain configurations on a $1001 \times 1001$ frequency grid spanning 2~THz. We reach this sample count with a 
GPU-accelerated implementation.

Figure~\ref{spectrum_PRL} presents the outcome, with simulated spectra in Fig.~\ref{spectrum_PRL}(a) and the measured dataset from Ref.~\cite{Smallwood2021} in Fig.~\ref{spectrum_PRL}(b). Because $B_\alpha$ drops sharply for the most strongly strained centers, the photoluminescence spectrum keeps only the weakly strained population near the unstrained resonance, recovering the narrow four-peak structure seen in the measured photoluminescence spectrum~\cite{Smallwood2021}. By contrast, the heterodyne spectrum [black trace, lower panel of Fig.~\ref{spectrum_PRL}(a)] retains the full inhomogeneous width.

A curious feature of the results is that the branching ratio described by Eq.~\eqref{branchingratio} needs to be exceptionally restrictive to accurately reproduce the experimental dataset depicted in Fig.~\ref{spectrum_PRL}(b), with a value of $\Gamma_\text{nr} / \Gamma_{\text{rad},e'} \gtrsim 10^6$ for strain values exceeding the filter threshold. The analysis conducted in Ref.~\cite{Smallwood2021} had not previously brought this feature to light. If the high-strain ratio $\Gamma_\text{nr} / \Gamma_{\text{rad},e'}$ is set much smaller, the heterodyne detected background is not sufficiently suppressed and begins to become visible in the simulation output. The origin of the effect lies in the fact that the strain tensor distribution is a multidimensional Gaussian, so the unfiltered fraction of \ce{SiV^-} centers scales down very quickly as strain threshold values are reduced. The excess population of strained and spectroscopically suppressed \ce{SiV^-} centers is weak on a center-by-center basis compared to unstrained centers, but becomes strong overall because strained color centers greatly outnumber unstrained ones.

In terms of experimental parameters, a ratio of $\Gamma_\text{nr} / \Gamma_{\text{rad},e'} \gtrsim 10^6$ carries a specific implication. If the nonradiative decay rate is on the order of 1 GHz (a lower limit set by the measured $\ce{SiV^-}$ ensemble homogeneous linewidth~\cite{Smallwood2021}), then the radiative linewidth could be no greater than 1 kHz. This corresponds to a minimum radiative lifetime of $1/(2\pi \Gamma_{\text{rad},e'}) \approx 160$ $\upmu$s. This result may be unreasonable and could serve as an indicator of model incompleteness or incorrectness, which could suggest the need for further studies and analyses. A possible resolution to this issue may be that the branching should not be defined as a function of the multidimensional strain tensor, but rather as a function of the one-dimensional spectroscopic shift. Alternatively, there may be some other as-yet neglected mechanism. This second possibility has some experimental support in recent studies, which have shown that strain effects can become visible from one location on the sample to another, even when photoluminescence detection is the only experimental probe used \cite{Bates2021}. Additional studies have also shown that \ce{SiV^-} centers seem to preferentially form along host lattice grain boundaries \cite{Angell2024}. There may be relevant effects associated with these grain boundaries that are not presently taken into account.

\section{Discussion and Conclusions}
\label{sec:conclusions}

We have developed a quantitative model of strain in the \ce{SiV^-} system, connecting the heterodyne-detected and photoluminescence-detected MDCS spectra reported in Ref.~\cite{Smallwood2021}. Building the single-center lineshape from density-matrix perturbation theory and averaging over an ensemble of strain configurations, we compared five strain distributions against the measured heterodyne spectrum. Fully random anisotropic strain offers the most consistent account of the observed broadening, while the uniaxial, uniaxial without transverse response, randomly oriented uniaxial, and isotropic distributions each retain a discrete peak structure that the measurements do not show. Reproducing the heterodyne lineshape requires shear strain alongside the dominant axial strain, with Gaussian widths near $2.8\times10^{-4}$ and $3.5\times10^{-5}$ for the axial and shear components and no systematic Poisson-ratio relationship between them.

The photoluminescence spectrum follows from the same strain environment once we include the branching ratio. Strain breaks the symmetry that forbids coupling to a dark state and opens a nonradiative decay channel in strongly strained centers. Because photoluminescence detection responds only to radiative decay, this channel removes strongly strained centers from the spectrum. Photoluminescence detection therefore isolates the weakly strained population near the unstrained resonance, while heterodyne detection retains the full inhomogeneous width. The same strain distribution that broadens the heterodyne spectrum, combined with a threshold on the largest strain component, reproduces the narrow photoluminescence spectrum and accounts for the contrast between the two detection schemes.

These results carry three broader implications. First, comparing heterodyne and photoluminescence MDCS separates radiative from nonradiative dynamics and exposes states that neither scheme resolves on its own, providing a general strategy for identifying dark states in other emitter systems. Second, the strong strain sensitivity of the \ce{SiV^-} spectrum suggests a route to non-contact strain sensing in diamond. Third, the same strain coupling that opens the nonradiative channel offers a handle for modulating radiative emission in \ce{SiV^-} devices. Direct confirmation of the dark state and its microscopic origin remains an open question, as does measurement of the nonradiative rate beyond the step approximation used here and characterization of how the branching ratio depends on temperature and the applied field.

\begin{acknowledgments}
We thank S.\ T.\ Cundiff and G.\ Thiering for useful discussions.
This material is based upon the work supported by the NSF under Grant Nos.\ DMR-2003493 and OAC-1626645.
\end{acknowledgments}

\section*{Data Availability}

Data that support the findings of this study were experimentally collected and also generated by numerical simulations. Datasets and source code and parameters used to generate the simulations are publicly available \cite{Straintheorydata2026}.

\bibliography{References}

\end{document}